\documentstyle{article}
\begin{document}
\begin{center}
{\LARGE\bf A quantum key distribution protocol without classical
communications}
\end{center}

\begin{center}
{Xiaoyu Li\\Institute of Computing Technology, Chinese Academy of Sciences\\
P.O.Box 2704, Beijing, 100080, P.R.China}
\end{center}

\begin{abstract}
We provide a quantum key distribution protocol based on the
correlations of the Greenburger-Horne-Zeilinger(GHZ) state.
 No classical communication is needed in the process of the
establishment of the key. Our protocol is useful when an
unjammable classical communication channel is unavailable. We
prove that the protocol is secure.
\end{abstract}

\section{Introduction}
Cryptography is the art of transmitting secret information over
insecure channels. If the two parties share a random sequence of
bits (a key), they can encrypt the plain text with the key. So how
to distribute a secret key is crucial for secure communication. In
classical cryptography it is the most difficult problem.

Quantum key distribution (QKD) protocol is a good way to solve
this problem in which the key is kept secret by the fundamental
principle of quantum mechanics. So it's secure. The first quantum
key distribution protocol is proposed by by C. H. Bennett and G.
Brassard in 1984 (so called BB84 protocol) [1]. Since then several
quantum key distribution protocols have been established and their
securities have been studied [2-6]. Many experiments have been
done [7-11]. In those protocols above there are two channels
needed: a quantum channel to transmit quantum qubits and a
classical channel to exchange classical information. Moreover the
classical channel must be unjammable though it could be insecure.
But it seems that this condition is not always practicable. If
there is no unjammable classical channel available, none of them
can succeed.

 In this paper we propose a new QKD scheme in which
classical communication isn't needed. It's based on the
correlations of GHZ state. Side A creates n three qubits system in
GHZ state and sends the third qubit of them to side B. Side B
encodes his key in the qubits and sends them back. When side A
receives the qubits, he can extract the key by performing
operation and measurement on the three qubits system. The delicate
nature of entanglement guarantees that no other people can get the
key.

The paper is organized as follows. In section 2 we briefly
introduce the basic idea on which our QKD protocol is based. Then
the protocol is provided in section 3. Next in section 4 we prove
that it is secure. At last we give a brief discussion on it.

\section{Basic Idea}
First we introduce the concept of the Bell state measurement
briefly. It is a measurement on a two-qubit system in which the
basis for measurement is the maximally entangled states
\begin{equation}
\begin{array}{lll}
|\Phi^{+}>={1\over \sqrt{2}}(|00>+|11>)\\
|\Phi^{-}>={1\over \sqrt{2}}(|00>-|11>)\\
|\Psi^{+}>={1\over \sqrt{2}}(|01>+|10>)\\
|\Psi^{-}>={1\over \sqrt{2}}(|01>-|10>).
\end{array}
\end{equation}
The four states are also called Bell states.

The GHZ state is proposed  by Greenburger, Horne and Zeilinger
[12]. It's an entangled state of a three qubits system. It is
expressed as
\begin{equation}
|\Delta>={1\over \sqrt{2}} (|000>+|111>).
\end{equation}
For simplicity we call the three-qubit system the tripartite
system. If we perform a CONTROLLED-NOT (CNOT) operation on the
first qubit (as the target qubit) and the second qubit (as the
control qubit). The state of the tripartite system is transformed
into
\begin{equation}
{1\over \sqrt{2}}(|000>+|011>)=|0>\otimes {1\over \sqrt{2}}
(|00>+|11>).
\end{equation}
So the first qubit is no longer entangled with the other two. The
tripartite system has been divided into two independent
subsystems. Consider the subsystem of the last two qubits, we will
notice that it is just in a Bell state $|\Phi^{+}>$.

On the other hand, If we perform a unitary operation $\sigma_x$ on
the third qubit of the GHZ state where
\begin{equation}
\sigma_x= \left(
\begin{array}{lcr}
0    &    1\\
1    &    0
\end{array}
\right).
\end{equation}
 Then we get
\begin{equation}
|\Delta^{'}>={1\over \sqrt{2}} (|001>+|110>)
\end{equation}
Performing the CNOT operation as above, we obtain
\begin{equation}
{1\over \sqrt{2}} (|001>+|010>)= |0>\otimes {1\over \sqrt{2}}
(|01>+|10>).
\end{equation}
So the subsystem of the last two qubits is also in a Bell state
$|\Psi^{+}>$. Obviously if we can do the Bell state measurement on
the last two qubits, we can distinguish the state equation(3) and
the state equation(6), that is to say, we can distinguish
$|\Delta>$ and $|\Delta^{'}>$ by the procedure above. So we can
apply this fact to establish a key shared by two sides.

\section{Quantum distribution protocol without classical
communications}

We assume that the two sides who want to communicate are Alice and
Bob. Eve, another one, wants to eavesdrop. If Alice wants to
communicate with Bob, they do as follows.

\noindent (1):Alice creates n tripartite systems in state $|\Delta
>$ and
sends the third qubit of each tripartite system to Bob.\\
(2):When Bob receives the qubits, he performs $\sigma_x$ operation
or does nothing at random on each qubit. When he performs
$\sigma_x$, he writes down '1' while when he does nothing, he
writes down '0'.
Finally he has a n-bit string {\sl k}.\\
(3):Bob sends the qubits back to Alice. \\
(4):After receiving every qubit, Alice combines it with the
corresponding two qubits in her hand. Then to every tripartite
system Alice performs CNOT operation on the first and the second
qubit in which the first one is the target qubit and the second
one is the control qubit. Then she does a Bell state measurement
on the last two qubits. When the measurement outcome is
$|\Phi^{+}>$, she writes down '0' and when it is $|\Psi^{+}>$, she
writes down '1'. But if the measurement outcome is $|\Phi^{-}>$ or
$|\Psi^{-}>$, the protocol fails. So Alice abandons it
and turns to (1). \\
(5):At last Alice also gets a n-bit string $k^{'}$ and she can
assure $k^{'}=k$. It is the key shared by both the two sides.

So far Alice and Bob have shared a key. The process of quantum key
distribution is completed. We can notice that no classical
communication is needed in the establishment of the key.

Finally there is a problem left. How does Bob know wether the key
has been established or not? Because there is no classical channel
available, Alice can't inform Bob by sending classical message as
in the previous protocols. We can stipulate that if Alice finds
that the protocol fails in step (4), she should turn to step (1)
and resend. When Bob receives qubits again, he knows that the
previous work has failed. So Bob begins to establish a new key. We
definite a maximal waiting time $t_c$. If Bob hasn't received the
qubits resent from Alice after $t_c$, he knows that the key has
been established successfully. On the other hand we stipulate the
terminal condition. If Alice has sent qubits to Bob for 10 times,
she still can't get the correct key. So she gives up and sends n
random qubits to Bob as the terminal signal. When Bob receives the
qubits from Alice for the 11st time, he knows that the effort to
establish a key from Alice has been terminated. So he also gives
up.

\section{Security of the quantum key distribution protocol}
This protocol is secure because of the fundamental principle in
quantum mechanics. If Eve intercepts and captures the qubits which
Bob sends to Alice, she can't prevent her existence from being
found. We give the proof of it below.

Now let's consider the possible attacks to the protocol. The aim
of eavesdropper, Eve, is to obtain the key.

First we can note that Eve can intercept and capture the qubits
when Alice sends them to Bob and when Bob sends them back to
Alice. But it's unhelpful for Eve to catch the qubits sent from
Alice to Bob. Since in fact the key is created by Bob, there is no
information about the key contained in the qubits which Alice
sends to Bob because the key hasn't been created. Actually from
our protocol all people including Eve know that each of these
qubits is the third qubit of a GHZ state $|\Delta>$. But this fact
is valueless for any eavedropper. So Eve will concentrate on the
qubits sent back from Bob to Alice.

 When Bob sends the
qubits back to Alice, Eve can catch the qubits. But measuring the
qubits is unhelpful, because the density matrix of the qubits is
\begin{equation}
\begin{array}{lll}
\rho & = & {\rm Tr} |\Delta > \\
& = &{{\rm Tr}}_{12} \{{1\over\sqrt{2}}
(|000>+|111>){1\over\sqrt{2}}(<000|+<111|)\}\\
& = & {1\over 2}(|0>+|1>)(<0|+<1|)
\end{array}
\end{equation}
or
\begin{equation}
\begin{array}{lll}
\rho & = & {\rm Tr} |\Delta^{'}>\\
 & = & {\rm Tr}_{12}
\{{1\over\sqrt{2}}
(|001>+|110>){1\over\sqrt{2}}(<001|+<110|)\}\\
& = & {1\over 2}(|0>+|1>)(<0|+<1|)
\end{array}
\end{equation}
where ${\rm Tr}_{12}$ denotes the partial trace over qubit 1 and
qubit 2. If she measures the qubits, she will get outcome '0' and
'1' with equal probability ${1\over 2}$. So she can't find wether
the state of the tripartite system is $|\Delta >$ or
$|\Delta^{'}>$, in other words, she can't get any information
about the key. Moreover Alice will find Eve's existence easily
when she gets the qubits sent back.

Consider the qubit on which Bob does nothing, the state of the
tripartite system is $|\Delta >$. If Eve's measurement outcome is
'0', the state of the tripartite system will be $|000>$. According
to the protocol Alice performs CNOT operation,
\begin{equation}
{\rm CNOT}:|000>\longrightarrow |000>.
\end{equation}
It can be noted that
\begin{equation}
|000>=|0>\otimes {1\over\sqrt{2}}(|\Phi^{+}>+|\Phi^{-}>).
\end{equation}
When Alice does the Bell state measurement on the last two qubits,
the outcomes is $|\Phi^{+}>$ and $|\Phi^{-}>$ with equal
probability ${1\over 2}$.
 If Alice's measurement outcome is '1', the state of
the tripartite system will be $|111>$. Then Alice performs CNOT
operation.
\begin{equation}
{\rm CNOT}:|111>\longrightarrow |011>.
\end{equation}
We can see
\begin{equation}
|011>=|0>\otimes {1\over\sqrt{2}} (|\Phi^{+}>-|\Phi^{-}>).
\end{equation}
Similarly Alice will get $|\Phi^{+}>$ and $|\Phi^{-}>$ with equal
probability ${1\over 2}$ by doing a Bell state measurement. But in
the protocol we know that if no eavesdroppers exist, it's
impossible for Alice to obtain measurement outcomes $|\Phi^{-}>$.
Once Alice gets $|\Phi^{-}>$, she knows that someone has
eavesdropped.

Then consider the qubit on which Bob does $\sigma_x$ operation,
the state of the tripartite system is $|\Delta^{'}>$. Eve catches
and measures the qubit sent back from Bob to Alice. When her
measurement outcome is '1', the state of the tripartite system is
transformed into $|001>$. After receiving the qubit, Alice
performs a CNOT operation
\begin{equation}
{\rm CNOT}:|001>\longrightarrow |001>.
\end{equation}
Doing the same reasoning as above,
\begin{equation}
|001>  =  |0>\otimes {1\over\sqrt{2}} (|\Psi^{+}>+|\Psi^{-}>).
\end{equation}
Performing the Bell state measurement, Alice will get $|\Psi^{+}>$
and $|\Psi^{-}>$ with equal probability ${1\over 2}$. When Eve's
measurement is '0', the state of the tripartite system is $|110>$.
Then Alice performs the CNOT operation
\begin{equation}
{\rm CNOT}:|110>\longrightarrow |010>.
\end{equation}
Similarly we know
\begin{equation}
|010>  =  |0>\otimes {1\over\sqrt{2}}(|\Psi^{+}>-|\Psi^{-}>).
\end{equation}
Obviously Alice gets $|\Psi^{+}>$ and $|\Psi^{-}>$ with equal
probability ${1\over 2}$ by doing the Bell state measurement. So
if Alice gets $|\Psi^{-}>$, she can assert the existence of Eve.

Then we calculate the probability that Eve escapes from being
found. We assume that Bob does $\sigma_x$ operation and does
nothing with equal probability ${1\over 2}$. The probability that
Alice gets correct measurement for one qubit though Eve has
attacked is
\begin{equation}
{1\over 2}\times {1\over 2}+{1\over 2}\times {1\over 2}
 ={1\over 2}.
\end{equation}
If n=1000, the probability that Alice couldn't find Eve's
existence after the establishment of the key has been completed is
\begin{equation}
P_{error}=({1\over 2})^{1000}\approx (10)^{-300}
\end{equation}
It's a digit too small to imagine. In fact we can say that it is
impossible for Eve to cheat without being found.
 So we can conclude that if Eve catches the qubits and
measures them, she can't prevent herself from being found, in
other words, the attack fails.

Now, we discuss another strategy of attack. Suppose that Eve
captures the qubit from Bob to Alice, she perform a CNOT
operation. The control qubit is the qubit captured and the target
qubit is an auxiliary qubit $|0>_E$ owned by Eve. After the CNOT
operation, the state of the four qubits system is
\begin{eqnarray}
\begin{array}{lll}
{\rm CNOT}:|\Delta >|0>_E \longrightarrow
  {1\over\sqrt{2}} & ( & |000>|0>_E\\
 & + & |111>|1>_E)
\end{array}
\end{eqnarray}
or
\begin{eqnarray}
\begin{array}{lll}
{\rm CNOT}:|\Delta^{'}>|0>_E \longrightarrow {1\over\sqrt{2}}
& ( & |001>|1>_E\\
& + & |110>|0>_E).
\end{array}
\end{eqnarray}
Alice performs the CNOT operation on the first and the second
qubit according the protocol, the state is transformed into
\begin{equation}
|S>_{Four}={1\over\sqrt{2}}|000>|0>_E+|011>|1>_E
\end{equation}
or
\begin{equation}
|S^{'}>_{Four}={1\over\sqrt{2}}|001>|1>_E+|010>|0>_E.
\end{equation}
Equation(21) can be rewritten as
\begin{eqnarray}
\begin{array}{lll}
|S>_{Four} & = & {1\over2}|0>(|\Phi^{+}>+|\Phi^{-}>)|0>_E\\
& + & {1\over2}|0>(|\Phi^{+}>-|\Phi^{-}>)|1>_E.
\end{array}
\end{eqnarray}
So when Alice does the Bell state measurement on the second and
the third qubit, she will obtain $|\Phi^{+}>$ and $|\Phi^{-}>$
with equal probability ${1\over 2}$. Equation(22) can be rewritten
as
\begin{eqnarray}
\begin{array}{lll}
|S^{'}>_{Four} & = & {1\over2}|0>(|\Psi^{+}>+|\Psi^{-}>)|1>_E\\
& + & {1\over2}|0>(|\Psi^{+}>-|\Psi^{-}>)|0>_E.
\end{array}
\end{eqnarray}
So Alice will get $|\Psi^{+}>$ and $|\Psi^{-}>$ with equal
probability ${1\over 2}$ by doing the Bell state measurement.

Similarly we can calculate the probability that Eve prevents her
from being found. The probability for one qubit is
\begin{equation}
{1\over 2}\times {1\over 2}+{1\over 2}\times {1\over 2}
 ={1\over 2}.
\end{equation}
If n=1000, the probability that Eve's cheating isn't found will be
\begin{equation}
P_{error}=({1\over 2})^{1000}\approx (10)^{-300}.
\end{equation}
So this strategy of attack also fails.

Now we have proved that our protocol is secure.

\section{Discussion and conclusion}
We provide a quantum key distribution protocol without classical
communication. In contrast to previous protocols it doesn't need
an auxiliary classical channel to complete the key distribution.
So it may be useful while a reliable classical channel is
unavailable. One disadvantage of this protocol is that it needs to
do operations on the qubits, such as $\sigma_x$ and CNOT. This
adds difficulties to carrying it out technically. But we can
expect the difficulties will be overcome by the development of
technology. Finally the protocol is proved to be secure.

\section*{acknowledgement}
 Thank academician of Chinese Academy of
Sciences Ruqian Lu for directing me into research.

\end{document}